\documentstyle[psfig]{menu95}
\begin{document}
\title{PHOTO- AND ELECTROPRODUCTION OF THE ETA AND ETA-PRIME
MESONS$^{*}$ \\
}
\author{Nimai C. Mukhopadhyay, J. -F. Zhang \\
{\it  Physics Department, Rensselaer Polytechnic Institute} \\
{\it Troy, NY 12180-3590, USA}
\and
 M. Benmerrouche \\
{\it Saskatchewan Accelerator Laboratory, University of
Saskatchewan}\\{\it Saskatoon, SK S7N 0W0 Canada }}
\maketitle

\begin{abstract}
We discuss the eta and eta prime photo- and electroproduction as
a way to probe the nucleon resonance structure.
\end{abstract}

\section{Introduction}

The Blaubeuren symposium of summer, 1995 has been a gathering
bathed in the excitement generated by the eta physics. We
have had two plenary sessions dedicated to this subject and
many fine contributions. Of particular interest has been the recent
experiments done at Mainz\cite{1} and Bonn\cite{2} facilities.
The Mainz photoproduction data\cite{1} are very precise.

On the theory side, Benmerrouche and Mukhopadhyay\cite{3} laid
out some time ago an effective Lagrangian approach(ELA), which emphasizes the
tree-level structure of the photoproduction reaction
\begin{equation}
\gamma +p\rightarrow p+\eta ,
\end{equation}
clearly dominated by the excitation of the N$^*$(1535)
resonance. Our recent work\cite{4} has dealt with all then available
data, most of which are in the higher energy region(E$_{\gamma}
\ge 900$ MeV). The publication of the precise Mainz data\cite{1},
along with the less precise ones from Bates\cite{5} and Bonn\cite{6}, has
necessitated a revisit of our ELA. This is what our report here
is about.

The electroproduction of eta
\begin{equation}
e+p\rightarrow e^{\prime}+p+\eta
\end{equation}
involves older data base\cite{7}. Our result essentially coincides
with  the thesis of Benmerrouche\cite{8}. We have not been
able to reproduce the non-isotropic angular distributions reported at
low $Q^2$ from Bonn\cite{9}. These data must be confirmed.
New data are urgently needed for
the reaction (2).

Our work on the theory of eta prime photoproduction, just about to
come out in print\cite{10}, is the first theoretical paper on this
subject, based on the old data base dating back to 1968\cite{11}.
As such, it is exploratory in nature. There is no experimental
work to guide us into the domain of eta prime electroproduction and we
shall not discuss that any further. There exists an approved
proposal on eta prime photoproduction at CEBAF\cite{12}.

We note with anticipation new theoretical efforts\cite{13}
to build on the coupled-channel analysis by Tanabe and
Bennhold\cite{14}. Confusions about the disputed data base on
the strong sector, discussed at this conference in a hot debate between
Bennhold\cite{13} and Svarc\cite{13}, do not allow us to draw any firm
conclusions on the unitarity and coupled-channel effects, as yet,
contrary claims notwithstanding. This symposium has featured
these concerns in  spirited discussions.

\section{Basic issues}

    In the context of the $\eta$ and $\eta^{\prime}$
photo- and electroproduction, basic questions mirror those of
the pi-zero production. Some of these are:

(1) What are the nature of the meson-nucleon couplings?
How large are the meson-nucleon coupling ``constants"?

(2) Are there any ``low-energy" theorems near thresholds of
the eta and eta prime photoproduction?

(3) What are the key contributions at the tree level? How important are the
roles of the various N$^*$ excitations?

(4) How big are the form factor effects? What are the exact
forms of these form factors?

(5) How large are the effects from unitarity?

We discuss some of these issues here. Others are picked up
later.

On the eta-nucleon coupling constant $g_{\eta}$ for the
pseudoscalar meson-nucleon coupling, one can give rather
broad range of values, using the SU(3) or SU(6)$_W$ symmetry
arguments\cite{17}. These bands imply $g_{\eta}$
considerably smaller than the value
of $g_{\pi}$. Using quark model arguments, we can calculate
these ranges for $g_{\eta^{\prime}}$ as well. Similarly, the
t-channel vector meson exchanges can be constrained by using
suitable constraints on the vector meson-nucleon couplings
\cite{3,4,10}.

Another important  point is the handling of the spin-3/2
baryon resonances, which involves proper treatment of spin-3/2
 propagators\cite{15}, often poorly done in the literature.
In view of our recent discussion of it\cite{4}, we shall not
elaborate this any further. Suffice to say that it
gives contribution in the spin-1/2 sector, which can be quite
important, but is often neglected. We also note that
no comparable treatments are
available in the literature on the higher spin baryon
reconances. Fortunately, their roles in the present context are
rather marginal.
\section{Our analysis of the Mainz photoproduction data}
The eta photoproduction data from Mainz obtained by Krusche {\it et al}. have
very small statistical and systematic errors. Our ELA fit to
this data yields excellent agreement (samples are in Fig.1). The same
fit agrees with the recent data of Dytman {\it et al}. from the Bates
laboratory\cite{5} at $E_{\gamma}=753$MeV, but disagrees
with the latter at $E_{\gamma}=729$MeV. Our tree-level ELA describes very well
 the total cross-section data of Mainz, Bonn\cite{6} and the Bates data
point at $E_{\gamma}=753$MeV.
\begin{figure}
\vspace{2.0in}
\centerline{\psfig{figure=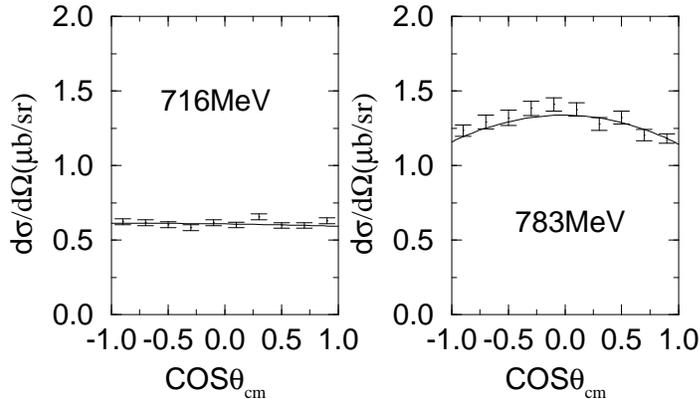,height=1.5in}}
\bigskip
\bigskip
\caption{ Our effective Lagrangian approach description of
the Mainz eta photoproduction data on proton\protect{\cite{1}}:
(a) $E_{\gamma}=716 MeV$ (b) $E_{\gamma}=783 MeV$. The main
parameter fitted here is $\xi$ of N$^*$(1535)[see text].}
\end{figure}

We can extract from the Mainz data an electrostrong parameter
characteristic of the N$^*$(1535) excitation and decay\cite{16}. This
quantity, defined by Benmerrouche and Mukhopadhyay\cite{3}, is
conservatively estimated to be
\begin{equation}
\xi =(2.20\pm0.15)\times 10^{-4} MeV^{-1}.
\end{equation}
Recent estimates in the quark model by Capstick and Roberts\cite{17}
has yielded a value of this parameter about half of the above, while Li,
in a recent preprint\cite{18}, extracts a figure within the
order of magnitude of Eq.(3). Clearly, this is a
fundamental baryon property that should be ultimately
computed reliably via QCD.

Other inferences from the Mainz data are as follows:

(1) We cannot determine precisely $g_{\eta NN}$ within our
ELA, given the relatively small role nucleon Born terms play. Our
extracted range is in fair agreement with SU(3).

(2) The vector meson sector compensates for the other non-
resonant contributions, thereby playing a useful role.

(3). The N$^*$(1520) [D$_{13}$] resonance also plays some role that we
can discern. The measured angular distributions at higher energies
($E_{\gamma}\ge 780$MeV) require the presence of this resonance. At energies
higher than $E_{\gamma}=900$MeV, other resonances become important\cite{4}.

\section{Our analysis of the data on electroproduction of
the etas}

  The data, rather old, show  generally flat angular distributions,
indicating the dominance of the N$^*$(1535) excitation. The data of
Brasse {\it et al}.\cite{7} continue to be flat for $Q^2\le 3 GeV^2$.

\section{Photoproduction of the eta prime meson}

The data come from the venerable ABBHHM collaboration\cite{11}
from 1968 and  are of poor quality. Our ELA analysis
\cite{10} is motivated by the quark model calculation of
Capstick and Roberts\cite{17}, who give us a first look at the
prediction of the electromagnetic excitation strength in the
quark model.

The physics of $\eta^{\prime}$ is interesting for many reasons:
$\eta_1$-$\eta_8$ mixing angle\cite{19}, chiral U(1) problem\cite{20},
the U(4)$\supset$SO(4) dynamical symmetry\cite{21}, just to mention a
few. We have found another wrinkle: the importance of s- and
u-channel form factors, without which the cross-section
violates unitarity very badly. The form factors we
use are:
\begin{equation}
F(s)=1/\left[ 1+\frac{(s-M_{R}^2 )^{2}}{\Lambda^4}
\right]
\end{equation}
where $\Lambda^2$ is of the order of 1$GeV^2$,
 $M_R$ is the resonance mass. Similar
forms can be used in the u-channel. Of course, we must respect gauge
invariance.

Our fit of the ABBHHM data (Fig.2) is quite satisfactory, given the quality
of the data. The surprising feature, about which we should
hopefully hear more in the next conference, is the important role of the
$D_{13}$ resonance, N$^*$(2080), which is strongly excited
in the quark model estimate by Capstick and Roberts.
\begin{figure}
\vspace{2.0in}
\centerline{\psfig{figure=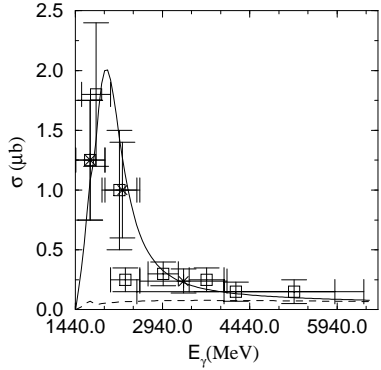,height=1.5in}}
\bigskip
\bigskip
\caption{Our ELA description of the eta prime photoproduction
data off protons\protect{\cite{11}} obtained by the ABBHHM
colleboration. The dashed line is the cross-section when
N$^*$(2080) is turned off.}
\end{figure}

\section{Concluding remarks}

The eta and eta prime photoproduction studies have ushered an exciting
frontier of the hadron physics  involving N$^*$ resonances. We
have obtained the first careful look at the  properties of
N$^*$(1535) from the precise Mainz data on photoproduction of eta mesons.
The photoproduction of eta prime promises to yield new valuable information
on the electrostrong properties of the N$^*$(2080) resonance,
about which we know very little. New neutron data from Mainz\cite{22,23}
and polarization studies from Bonn for eta photoproduction, together
with the advent of CEBAF, should bode well for the future of this
field.

\section{Acknowledgements}

   One of us(NCM) is grateful to Prof. Gerhard J. Wagner  for his
enthusiastic invitation and support.
He is also grateful to Prof. Bernd Krusche for his
many helpful remarks and kind hospitality. This work is
supported at RPI by the U.S. Department of Energy and  at SAL by
the Natural Sciences and Engineering Research Council of
Canada.

$^{*}$ Invited talk presented by Nimai C. Mukhopadhyay at
the sixth International Symposium on the Meson-Nucleon Physics
and the Structure of the Nucleon, Blaubeuren-T\"{u}bingen,
Germany, July 1995.

\end{document}